\documentclass[a4paper, 12pt, onecolumn]{article}
\pdfoutput=1

\usepackage[cmex10]{amsmath}	
\usepackage{amssymb}

\usepackage{mathtools}	
\mathtoolsset{showonlyrefs}

\usepackage[T1]{fontenc}
\usepackage[latin1]{inputenc}
\usepackage{lmodern}

\newcommand{\RR}{\ensuremath{\mathbb{R}}}

\renewcommand{\P}{\ensuremath{\mathcal{P}}}	
\renewcommand{\S}{\ensuremath{\mathcal{S}}}	
\newcommand{\X}{\ensuremath{\mathcal{X}}}

\newcommand{\st}{\ensuremath{\mathop{\mathrm{st}}}}
\newcommand{\sq}{\ensuremath{\mathop{\mathrm{sq}}}}
\newcommand{\pd}[2]{\ensuremath{\frac{\partial #1}{\partial #2}}}
\newcommand{\diag}{\ensuremath{\mathop{\mathrm{diag}}}}

\newcommand{\vect}[1]{\ensuremath{\boldsymbol{#1}}}	
\newcommand{\TODO}[1]{}

\usepackage[nolist]{acronym}

\begin{acronym}
	\acro{PPM}{Prediction by Partial Matching}
	\acro{DMC}{Dynamic Markov Coding}
	\acro{CTW}{Context Tree Weighting}
	\acro{PAQ}{``Pack''}
	\acro{AC}{Arithmetic Coding}
	\acro{CM}{Context Mixing}
	\acro{CG}{Conjugate Gradient}
	\acro{KT}{Krichevsky-Trofimov}
	\acro{iid}{independent identically distributed}
	\acro{BWT}{Burrows-Wheeler-Transform}
	\acro{BFGS}{Broyden-Fletcher-Goldfab-Shanno}
	\acro{KKT}{Karush-Kuhn-Tucker}
	\acro{WFC}{Weighted Frequency Counting}
	\acro{MTF}{Move-to-Front}
	\acro{LP}{Laplace}
	\acro{SAKDC}{Swiss Army Knife Data Compression}
	\acro{SQP}{Sequential Quadratic Programming}
	\acro{bpc}{bits per character}
\end{acronym}

\usepackage[margins=normal,tracking=normal,leading=normal]{savetrees} 
\usepackage[top=1in, left=1.25in, textwidth=6in, textheight=9in]{geometry} 	
\title{\Large{Mixing Strategies in Data Compression}}
\author{
	\vspace{2mm}
	Christopher Mattern \\ 
	Fakultät für Informatik und Automatisierung\\
	Technische Universität Ilmenau\\
	Ilmenau, Germany \\
	\texttt{christopher.mattern@tu-ilmenau.de}
}
\date{}	

\begin{document}


\thispagestyle{empty}

\par\noindent\textbf{This paper is a preprint (IEEE ``accepted'' status).}

\bigskip\par\noindent
\textbf{IEEE copyright notice.}~\mbox{\copyright{}} 2012 IEEE. Personal use of this material is permitted. Permission from IEEE must be obtained for all other uses, in any current or future media, including reprinting/republishing this material for advertising or promotional purposes, creating new collective works, for resale or redistribution to servers or lists, or reuse of any copyrighted component of this work in other works.	
\bigskip\par\noindent
\textbf{DOI.}  10.1109/DCC.2012.40
\bigskip\par\noindent
http://doi.ieeecomputersociety.org/10.1109/DCC.2012.40
\newpage

\maketitle
\thispagestyle{empty}	



\begin{abstract}
\noindent
We propose geometric weighting as a novel method to combine multiple models in data compression. Our results reveal the rationale behind \acs{PAQ}-weighting and generalize it to a non-binary alphabet. Based on a similar technique we present a new, generic linear mixture technique. All novel mixture techniques rely on given weight vectors. We consider the problem of finding optimal weights and show that the weight optimization leads to a strictly convex (and thus, good-natured) optimization problem. Finally, an experimental evaluation compares the two presented mixture techniques for a binary alphabet. The results indicate that geometric weighting is superior to linear weighting.
\end{abstract}


\section{Introduction} \label{sec:introduction}

\subsection{Background} \label{sec:introduction_background}
The combination of multiple models is a central aspect of many modern data compression algorithms, such as \ac{PPM} \cite{thesis_bunton, hbdc, ppm_ii}, \ac{CTW} \cite{ctw_extensions, ctw_95} or \ac{PAQ} \cite{cm_paq6, hbdc}. All of these algorithms belong to the class of statistical data compression algorithms, which share a common structure: The compressor consists of a \emph{model} and a \emph{coder}; and it processes the data (a string $x^n\in\X^n$ for some alphabet $\X, |\X|\geq 2$) sequentially. In the $k$-th step, $1\leq k\leq n$, the model estimates the probability distribution $P( ~\cdot \mid x^{k-1})$ of the next symbol based on the already processed sequence $x^{k-1}=x_1 x_2 \dots x_{k-1}$. The task of the coder is to map a symbol $x\in\X$ to a codeword of a length close to $-\log P(x\mid x^{k-1})$ bits (throughout this paper $\log$ is to the base two). For decompression the coder maps the encoding, given $P(~\cdot\mid x^{k-1})$, to $x$. \ac{AC} closely approximates the ideal code length and is known to be asymptotically optimal \cite{eoit}. Therefore, the prediction accuracy of the model is crucial for compression.

Mixture models or mixtures combine multiple models into a single model suitable for encoding. Let us now consider a simple example, which gives two reasons for our interest in mixtures. First, assume that we have $m>1$ models available. Model $i, 1\leq i\leq m$, maps an arbitrary $x^n$ to a prediction $P_i(x^n)$ (a probability distribution), where
\begin{equation}
	P_i(x^n) = \prod_{k=1}^n P_i(x_k\mid x^{k-1}) = \prod_{k=1}^n \frac{P_i(x^k)}{P_i(x^{k-1})} \label{eq:compute_mixture_sequentially}
\end{equation}
and $P_i(x^k)>0, ~1\leq i\leq m, ~k\geq 0$.
When we compress $x^n$ with a single model $i$, we need to encode the choice of $i$ in $-\log W(i)$ bits (where $W(i)$ is the prior probability of selecting model $i$) and we need to store the encoded string, which adds $-\log P_i(x^n)$ bits. If we knew $x^n$ in advance, we could select 
\begin{equation}
	i = \arg \min_{1\leq j\leq m} \left[-\log(W(j)) - \log(P_j(x^n)) \right].	\label{eq:two_part_code}
\end{equation}
Surprisingly (as previously observed in e.g., \cite{universal_prediction}), a simple linear mixture $P(x^n) := \sum_{j=1}^m W(j) P_j(x^n)$ will never do worse than \eqref{eq:two_part_code}, since
\begin{align}
	-\log(W(i)) - \log(P_i(x^n)) &= -\log(W(i)P_i(x^n)) \\
	 &\geq -\log \sum_{j=1}^m \left( W(j)P_j(x^n)\right),
\end{align}
where $i$ is the model that minimizes \eqref{eq:two_part_code}. Such a mixture makes it possible to combine the advantages of different models without cumulating their disadvantages. Secondly, the sequential processing allows us to refine the mixture adaptively (in favor of the locally more accurate models).


\subsection{Previous Work}
Most of the major statistical compression techniques (\ac{PPM}, \ac{CTW} and \ac{PAQ}) are based on mixtures. In \ac{PPM} the concept of ``escape'' symbols is related to the computation of a recursively defined mixture distribution. The escape probability plays the role of a weight in a linear mixture. In \cite{thesis_bunton} Bunton gave a very comprehensive (at that time) synopsis on that topic. Previously, several different methods for the estimation of escape probabilities had been proposed, e.g., PPMA, PPMB, PPMC, PPMD, PPMP, PPMX \cite{hbdc}, PPMII \cite{ppm_ii}. \ac{CTW} relies on the efficient combination of exponentially many (depending on a ``tree depth'' parameter) models for tree sources. However, the structure of \ac{PPM} and \ac{CTW} restrict the type of models they combine (order-$N$ models for \ac{PPM} and models for tree sources for \ac{CTW}).
Recently, some of the techniques of \ac{CTW} led to $\beta$-weighting \cite{cm_cmidc}, as a linear general-purpose weighting method. We are interested in general-purpose mixture techniques, which combine arbitrary (and eventually totally different) models. The practical success of this approach was initiated by Mahoney with \ac{PAQ} (see \cite{hbdc} for details). \ac{PAQ} combines a large amount of totally different models (e.g., models for text, for images, etc.). As a minor part earlier work we successfully employed a simple linear mixture model for encoding \ac{BWT} output and proposed a method for the parameter optimization on training data \cite{cm_ccp2011}.


\subsection{Our Contribution}
In Section \ref{sec:geometric_model} we propose geometric weighting as a novel non-linear mixture technique. We obtain the geometric mixture as the solution of a divergence minimization problem. In addition we show that \ac{PAQ}-mixing is a special case of geometric weighting for a binary alphabet. Since geometric weighting depends on a set of weights, we examine the problem of weight optimization and propose a corresponding optimization method. In Section \ref{sec:linear_model} we focus on linear mixtures. In a fashion analogous to Section \ref{sec:geometric_model} we describe a new generic linear mixture and investigate the problem of weight optimization. Finally, we compare the behavior of the implementations (for a binary alphabet) of the two proposed mixture techniques and of $\beta$-weighting in Section \ref{sec:experimental_evaluation}.  Results indicate that geometric weighting is superior to the other mixture methods.


\section{Preliminaries} \label{sec:preliminaries}

First, we fix some notation. Let $\X$ denote an alphabet of cardinality $1<|\X|<\infty$ and let $x_i^j = x_i x_{i+1} \dots x_j$ be a sequence of length $n=j-i+1$ over $\X$. For short we may write $x^n$ for $x_1^n$. Abbreviations such as $(a_i)_{1\leq i\leq n}$ expand to $(a_1~a_2~\dots~a_n)$ and denote row vectors. Boldface letters indicate matrices or vectors, ``${}^T$'' denotes the transpose operator, $\vect 1_m := (1~1~\dots~1)^T\in \RR^m$ and $\Omega_m := \{\vect v\in\RR^m\mid \vect v\geq\vect 0,~\vect v^T \vect 1_m = 1\}$. We use $\log$ to denote the logarithm with base two, $\ln$ denotes the natural logarithm.

Suppose that we want to compress a string $x^n\in\X^n$ sequentially. In every step $1\leq k\leq n$ a \emph{model} $M:\cup_{k\geq 0} \X^k\rightarrow\P$ maps the already known prefix $x^{k-1}$ of $x^n$ to a \emph{model distribution} $P(~\cdot\mid x^{k-1}),~ P\in\P$, where $\P:=\{ Q:\X \rightarrow (0, 1) \mid \sum_{x\in\X} Q(x) = 1 \}$. An encoder translates this into a code of length close to $-\log P(x\mid x^{k-1})$ bits for $x$. Now, if there are $m>1$ \emph{submodels} $M_1, M_2, \dots, M_m$ (or submodels $1, 2, \dots, m$, for short), we require a \emph{mixture function} $f_k\colon\X\times \P^m\rightarrow(0, 1)$ to map the $m$ corresponding distributions $P_1, P_2, \dots, P_m$ to a single distribution $P(x)=f_k(x,P_1, P_2, \dots, P_m), P\in\P$, in step $k$; $f_k$ may depend on $x^{k-1}$.

An approach in information theory is to suppose that $x^n$ was generated by an unknown mechanism, which is called a \emph{source}. W.l.o.g. we may assume that $x^n$ was generated sequentially: In every step $k$ the source draws $x$ according to an arbitrary \emph{source distribution} \mbox{$P'\in \S := \{ Q:\X \rightarrow [0, 1] \mid \sum_{x\in\X} Q(x) = 1 \}$} (i.e., the distribution $P'$ may vary from step to step) and appends it to $x^{k-1}$ to yield $x^k = x^{k-1}x$. When we encode $x$, using a model distribution $P\in\P$, we obtain an expected code length of
\begin{equation}
	\sum_{x\in\X} P'(x) \log \frac{1}{P(x)} = \underbrace{\sum_{x\in\X} \left[ P'(x)\log \frac{1}{P'(x)} \right]}_{H(P')} + \underbrace{\sum_{x\in\X} P'(x)\left[ \log\left(\frac{1}{P(x)}\right) - \log \frac{1}{P'(x)} \right]}_{D(P'\parallel P)},
\end{equation}
where $H(P')$ is the \emph{source entropy} and $D(P'\parallel P)$ is the \emph{KL-divergence} \cite{eoit}, which measures the redundancy of $P$ relative to $P'$. Our aim is to find a $P$, that minimizes the code length. Since $H(P')$ is fixed (by the source), we want to minimize $D(P'\parallel P)$. We have $D(P'\parallel P) \geq 0$, which is zero iff $P=P'$, i.e., the best model distribution is the source distribution itself.


\section{Geometric Mixtures}	\label{sec:geometric_model}
This section contains the major part of our work: We derive geometric weighting as a novel method for combining multiple models. Now suppose that we have $m$ model distributions $P_1, P_2, \dots, P_m$ available in step $k$. Since the source distribution $P'$ is \emph{unknown} (if it exists at all) we try to identify an approximate source distribution $P\in\S\cap\P$, which we can use as a model distribution. It should be ``close'' (in the divergence-sense) to good models and ``far away'' from bad models. The terms good and bad refer to short and long code lengths (due to past observations and/or prior knowledge). We assume that we are given a set of non-negative weights $w_i, 1\leq i\leq m$, $\sum_{i=1}^m w_i>0$ (in Section \ref{sec:geometric_model_weight_estimation} we discuss a method of weight estimation), which quantify how well model $i$ fits the unknown source distribution. Summarizing, we are looking for the distribution
\begin{equation}
	P := \arg \min_{Q\in\P} \sum_{i=1}^m w_i D(Q \parallel P_i). \label{eq:geometric_model_target3}
\end{equation}


\subsection{Divergence Minimization} \label{sec:geometric_model_divergence_minimization}
In order to solve \eqref{eq:geometric_model_target3} we adopt the method of Lagrangian multipliers. First, we set $Q(x\mid x^{k-1}) = \theta_x$ and $\vect\theta^T = (\theta_x)_{x\in\X}$ to omit the implicit dependence on $k$ and to simplify the equations. Now we rewrite \eqref{eq:geometric_model_target3} to yield
\begin{align}
	\min_{\vect \theta} &\sum_{i=1}^m w_i \sum_{x\in\X} \theta_x \left[\log(\theta_x) - \log(P_i(x\mid x^{k-1})) \right] \label{eq:geometric_model_target4},\\
	\text{s.t.} & \sum_{x\in\X} \theta_x = 1 \text{ and } \theta_x > 0, x\in\X \nonumber
\end{align}
and formulate its Lagrangian
\begin{align}
	L(\vect \theta, \lambda, \vect \mu) =& \sum_{i=1}^m w_i \sum_{x\in\X} \theta_x \left[\log(\theta_x) - \log(P_i(x\mid x^{k-1})) \right] \nonumber \\
	 & - \lambda\left(1-\sum_{x\in\X} \theta_x\right) - \sum_{x\in\X} \left( \mu_x \theta_x \right).
\end{align}
The variable $\lambda$ and the vector $\vect \mu = (\mu_x)_{x\in\X}$ denote the Lagrange multipliers. A local minimum $\vect \theta^*, \lambda^*, \vect \mu^*$ satisfies the \ac{KKT} conditions (see, e.g. \cite{bertsekas})
\begin{gather}	
	\pd{L(\vect \theta^*, \lambda^*, \vect \mu^*)}{\theta_x} =  0, \label{eq:geometric_model_kkt1} \\	 
	\theta_x^* > 0,\ \mu_x^* \geq 0,\ \theta_x^* \mu_x^* = 0 \label{eq:geometric_model_kkt2}
\end{gather}
for all $x\in\X$ and
\begin{equation}
	\sum_{x\in\X} \theta_x^* = 1.	\label{eq:geometric_model_equality_constraint}
\end{equation}
Due to \eqref{eq:geometric_model_kkt2} we obtain $\mu_x^*=0$ for all $x\in\X$. Equation \eqref{eq:geometric_model_kkt1} can be transformed to
\begin{equation}
	\left(\sum_{i=1}^m w_i\right) + \lambda + \log\left(\theta_x^{* \sum_{i=1}^m w_i}\right) = \log \prod_{i=1}^m P_i(x\mid x^{k-1})^{w_i}. \label{eq:geometric_model_kkt1_simplified}
\end{equation}
Now we fix a disjoint pair $x\neq x'$ of symbols from $\X$ and subtract the corresponding instances of \eqref{eq:geometric_model_kkt1_simplified}, which results in
\begin{equation}
	\theta_{x'}^* = \theta_x^* \prod_{i=1}^m \left[ \frac{P_i(x'\mid x^{k-1})}{P_i(x\mid x^{k-1})} \right]^{w'_i}, \text{ where } w'_i := \frac{w_i}{\sum_{j=1}^m w_j}. \label{eq:geometric_model_theta_fraction}
\end{equation}
Again, we fix a single character $x$ and substitute any other occurrence of $x'\neq x$ in \eqref{eq:geometric_model_equality_constraint} via \eqref{eq:geometric_model_theta_fraction}. Thus we have
\begin{equation}
	1 = \theta_x^* + \theta_x^* \sum_{  x' \in \X \setminus \{x\}  } \prod_{i=1}^m \left[ \frac{P_i(x'\mid x^{k-1})}{P_i(x\mid x^{k-1})}\right]^{w'_i}, \label{eq:geometric_model_equality_constraint_simplified}
\end{equation}
which we rewrite to yield
\begin{equation}	
	\theta_x^* = \frac{ \prod_{i=1}^m P_i(x\mid x^{k-1})^{w'_i}}{ \sum_{x'\in\X} \prod_{i=1}^m P_i(x'\mid x^{k-1})^{w'_i}}. \label{eq:geometric_model_minimizer1}
\end{equation}
Finally, we reintroduce the dependencies on $k$ and obtain the geometric mixture 
\begin{equation}
	P(x\mid x^{k-1}) = f_k(x, P_1, P_2, \dots, P_m) := \frac{ \prod_{i=1}^m P_i(x\mid x^{k-1})^{w_i/(\vect w^T \vect 1_m)}}{ \sum_{x'\in\X} \prod_{i=1}^m P_i(x'\mid x^{k-1})^{w_i/(\vect w^T \vect 1_m)}}, \label{eq:geometric_model_mixture}
\end{equation}
where $\vect w^T = (w_i)_{1\leq i\leq m}$ is composed of the non-negative weights $w_i$. It remains to show that \eqref{eq:geometric_model_minimizer1} minimizes \eqref{eq:geometric_model_target4}. For this, we observe that the Hessian of \eqref{eq:geometric_model_target4} is
\begin{equation}
	\vect w^T \vect 1_m \cdot \diag\left( (1/\theta_x)_{x\in\X} \right),
\end{equation}
which is positive definite, since $\theta_x > 0$ for all $x\in\X$. 

\subsection{Weight Estimation and Convexity} \label{sec:geometric_model_weight_estimation}
The mixture function \eqref{eq:geometric_model_mixture} requires $m$ non-negative weights $\vect w^T = (w_i)_{1\leq i\leq m}$, which we still need to obtain.
In our situation the sequence $x^n$ is known (and fixed) and the sequence probability is given as a function of $\vect w$ as
\begin{align}
	\prod_{k=1}^n f_k(x_k, P_1, P_2, \dots, P_m) \label{eq:geometric_model_weight_likelihood} 
	= \prod_{k=1}^n  \frac{ \prod_{i=1}^m P_i(x_k\mid x^{k-1})^{w_i/(\vect w^T \vect 1_m)}}{\sum_{x'\in\X} \prod_{i=1}^m P_i(x'\mid x^{k-1})^{w_i/(\vect w^T \vect 1_m})}.
\end{align}
We now wish to find a weight vector $\vect w$, which maximizes \eqref{eq:geometric_model_weight_likelihood} (a maximum-likelihood estimation). Since a maximization of the sequence probability is equivalent to a minimization of its code length, we may alternatively solve
\begin{align}
	\min_{\vect w} \sum_{k=1}^n \left( -\log \frac{\prod_{i=1}^m P_i(x_k\mid x^{k-1})^{w_i/(\vect w^T \vect 1_m)}}{\sum_{x'\in\X} \prod_{i=1}^m P_i(x'\mid x^{k-1})^{w_i/(\vect w^T \vect 1_m)}} \right). \label{eq:geometric_model_weight_target1} 
\end{align}
We define $\vect w^*$ to be the minimizer of \eqref{eq:geometric_model_weight_target1}.

Now we want to show that the cost function of \eqref{eq:geometric_model_weight_target1} is convex. Since the cost function is a sum, we analyze a slight modification of a single term $l(\vect w) := -\ln(g(\vect w)/h(\vect w))$ (since $\log(x) \sim \ln(x)$). W.l.o.g. we may assume that $\vect w \in \Omega_m$ (due to \eqref{eq:geometric_model_theta_fraction}). In order to simplify the analysis of the Hessian of $l(\vect w)$ we set
\begin{align}
	g(\vect w) &:= \prod_{i=1}^m P_i(x_k\mid x^{k-1})^{w_i} = e^{\sum_{i=1}^m w_i \ln P_i(x_k\mid x^{k-1})} = e^{ \vect w^T \vect Q(x_k)}, \\
	h(\vect w) &:= \sum_{x\in\X} \prod_{i=1}^m P_i(x\mid x^{k-1})^{w_i} = \sum_{x\in\X} e^{\vect w^T \vect Q(x)}, \\
	\vect Q(x)^T &:= (\ln P_i(x\mid x^{k-1}))_{1\leq i\leq m}, \\
	p_x &:=\, e^{\vect w^T \vect Q(x)} / \sum_{x'\in\X} e^{\vect w^T \vect Q(x')} = f_k(x, P_1, P_2, \dots, P_m)
\end{align}
and we obtain
\begin{align}
	\nabla g(\vect w)/g(\vect w) &= \vect Q(x_k), & \nabla^2 g(\vect w)/g(\vect w) &= \vect Q(x_k) \vect Q(x_k)^T, \\
	\nabla h(\vect w)/h(\vect w) &= \sum_{x\in\X} p_x \vect Q(x), & \nabla^2 h(\vect w)/h(\vect w) &= \sum_{x\in\X} p_x \vect Q(x) \vect Q(x)^T.	
\end{align}
The Hessian of $l(\vect w)$ is positive definite, since for $\vect v \neq 0, \vect v \in \RR^m$
\begin{eqnarray}
	\vect v^T \nabla^2 l(\vect w) \vect v
	&=& \vect v^T\left( \frac{\nabla g(\vect w) \nabla g(\vect w)^T}{g(\vect w)^2} -
		\frac{\nabla^2 g(\vect w)}{g(\vect w)} +
		\frac{\nabla^2 h(\vect w)}{h(\vect w)} -
		\frac{\nabla h(\vect w) \nabla h(\vect w)^T}{h(\vect w)^2}  \right) \vect v \\	
	&=& \left( \frac{\vect v^T \nabla g(\vect w)}{g(\vect w)} \right)^2 -
		\vect v^T \frac{\nabla^2 g(\vect w)}{g(\vect w)} \vect v +
		\vect v^T \frac{\nabla^2 h(\vect w)}{h(\vect w)} \vect v -
		\left( \frac{\vect v^T \nabla h(\vect w)}{h(\vect w)} \right)^2 \\	
	&=& \left(\vect v^T \vect Q(x_k)\right)^2 - 
		\left(\vect v^T \vect Q(x_k)\right)^2 + 
		{\sum_{x\in\X}} \left( p_x \left(\vect v^T \vect Q(x)\right)^2 \right) -
		\left( \sum_{x\in\X} p_x \vect v^T \vect Q(x) \right)^2 \\
	&=& \sum_{x\in\X} \left( p_x \left(\vect v^T \vect Q(x)\right)^2 \right) -
		\left(\sum_{x\in\X} p_x \vect v^T \vect Q(x)\right)^2 \\		
	&>& 0
\end{eqnarray}
holds, where the last line is due to Jensen's inequality (since $\sum_{x\in\X}p_x=1$). It follows that the problem \eqref{eq:geometric_model_weight_target1} is strictly convex and there exists a single global minimizer $\vect w^*\in\Omega_m$.

We solve the problem \eqref{eq:geometric_model_weight_target1} with an optimization method tailored to a natural requirement in statistical compression: The sequence to be compressed is processed only once. Since the cost function is convex, the optimization algorithm does not need strong global search capabilities. A possible method-of-choice is an instance of iterative gradient descent \cite{bertsekas}. In the $k$-th step we use the estimates $\vect w(k)$ in place of $\vect w^*$ (in \eqref{eq:geometric_model_mixture}). Initially we set $\vect w(0) = 1/m \cdot \vect 1_m$. In each step $k$ we adjust the weight vector $\vect w(k-1)$ after we observe $x_k$ via a step towards the direction of steepest descent, i.e.,
\begin{equation}
	- \alpha_k \nabla_{\vect w} \left( -\log f_k(x_k, P_1, P_2, \dots, P_m) \right) \label{eq:iterative_gradient_descent}	
\end{equation}
where $\alpha_k>0$ is the step size in the $k$-th step. The choice of $\alpha_k$ is crucial for the convergence of $\vect w(k)$ to $\vect w^*$ \cite{bertsekas} (see Sections \ref{sec:geometric_model_paq} and \ref{sec:experimental_evaluation}). In the case of a geometric mixture function we have
\begin{align}
	\vect w(k) := \max\left\{\varepsilon\vect 1_m,~
		\vect w(k-1) + \alpha_k \frac{(\vect Q(x_k) - q_{x_k} \vect 1_m)
			- \sum_{x\in\X} p_x \left(\vect Q(x) - q_x\vect 1_m\right)}{\vect w^T \vect 1_m}\right\}, \label{eq:geometric_model_weight_update}
\end{align}
where $q_x := (\vect w^T \vect Q(x)) / (\vect w^T \vect 1_m)$. As an implementation detail $\varepsilon>0$ is a small constant to bound the weights away from zero and to avoid a division by zero in \eqref{eq:geometric_model_mixture}.


\subsection{PAQ Mixtures or Geometric Mixtures for a Binary Alphabet} \label{sec:geometric_model_paq}
Before we examine the details of ``the'' \ac{PAQ} mixture method, we need to clarify that there exist multiple \ac{PAQ} mixture mechanisms \cite{hbdc}. We focus on the latest instance, which was introduced in 2005 as a part of PAQ7. \ac{PAQ} computes mixtures for a binary alphabet and works with the probability of one-bits. The mixture is defined as follows
\begin{align}
	f_k(1, P_1, P_2, \dots, P_m) &:= \sq\left( \sum_{i=1}^m w_i(k-1) \st(P_i(1\mid x^{k-1})) \right), \label{eq:paq_mixture} \\
	w_i(k) &:= w_i(k-1) + \alpha (x_k-f_k(1,P_1, P_2, \dots, P_m)) \st(P_i(1)),\label{eq:paq_weight_update}
\end{align}
where $x_k$ is the bit we observed in step $k$ and
\begin{align}
	\st(x) := \ln \frac{x}{1-x},\quad\sq(x) := \frac{1}{1+e^{-x}}. \label{eq:paq_transforms}
\end{align}
Let $\vect w^T = (w_i)_{1\leq i\leq m}$ be the weight vector in step $k$ where we assume that $\vect w\in\Omega_m$. Now we rewrite \eqref{eq:paq_mixture} (due to \eqref{eq:paq_transforms}) to yield
\begin{align}
	f_k(1, P_1, P_2, \dots, P_m) 
	&= \left[1 + \exp\left(-\sum_{i=1}^m w_i \ln\frac{P_i(1\mid x^{k-1})}{1-P_i(1\mid x^{k-1})}\right) \right]^{-1} \label{eq:paq_mixture1} \\
	&= \left[1 + \prod_{i=1}^m \left(\frac{1-P_i(1\mid x^{k-1})}{P_i(1\mid x^{k-1})}\right)^{w_i} \right]^{-1} \\
	&= \frac{ \prod_{i=1}^m P_i(1\mid x^{k-1})^{w_i}}{ \prod_{i=1}^m P_i(0\mid x^{k-1})^{w_i} + \prod_{i=1}^m P_i(1\mid x^{k-1})^{w_i}},
\end{align}
which matches \eqref{eq:geometric_model_mixture}. It is easy to check (via substituting \eqref{eq:paq_mixture} into \eqref{eq:iterative_gradient_descent}), that \eqref{eq:paq_weight_update} is an instance of iterative gradient descent, where $\alpha_k=\alpha$ is constant in any step and the $\max$-operation is omitted.
When $\alpha$ is sufficiently small, the sequence $(\vect w(k))_{k\geq 1}$ converges to some $\vect w_\alpha$ rather than the optimal solution $\vect w^*$. In turn, $\lim_{\alpha\rightarrow 0} \vect w_\alpha = \vect w^*$ \cite{bertsekas}. A (small) constant step size $\alpha$ thus needs to be determined experimentally.


\section{Linear Mixtures} \label{sec:linear_model}

Let us return to the setting of Section \ref{sec:introduction_background}. Instead of encoding $x^n$ with model $i$ and transmitting our choice in $-\log W(i)$ bits, we will not do worse using the mixture distribution
\begin{equation}
	P(x^n) := \sum_{i=1}^m W(i) P_i(x^n).
\end{equation}
Since we want to process $x^n$ sequentially we use the distribution (cf. \eqref{eq:compute_mixture_sequentially})
\begin{align}
	 \frac{P(x^{k-1}x)}{P(x^{k-1})} 
	 &= \frac{\sum_{i=1}^m P_i(x^{k-1}x) W(i)}{P(x^{k-1})} \\
	 &= \sum_{i=1}^m \frac{P_i(x^{k-1})W(i)}{P(x^{k-1})} P_i(x\mid x^{k-1}) \\
	 &= \sum_{i=1}^m W(i\mid x^{k-1}) P_i(x\mid x^{k-1}) \label{eq:linear_model_mixture}
\end{align}
in step $k$. There is an obvious interpretation for the mixture \eqref{eq:linear_model_mixture}. Suppose that there are $m$ sources and a probabilistic switching mechanism, which selects source $i$ with probability $W(i \mid x^{k-1})$ in step $k$ (we interpret this as the posterior probability of $i$ given $x^{k-1}$). When a source is selected, it appends a character $x$ (with probability $P_i(x\mid x^{k-1})$) to the sequence $x^{k-1}$ to yield $x^k = x^{k-1} x$. We denote such a source as a \emph{switching source}.


\subsection{$\beta$-Weighting}
We can modify the probability assignment of \eqref{eq:linear_model_mixture} to yield a linear mixture technique called \emph{$\beta$-weighting}, which has its roots in the CTW compression technique and was proposed in \cite{cm_cmidc}. $\beta$-weighting is defined by
\begin{align}
	f_k(x, P_1, P_2, \dots, P_m) &:= \sum_{i=1}^m \beta_i(k) P_i(x\mid x^{k-1}), \label{eq:beta_weighting_mixture}\\
	\beta_i(k) &:= W(i\mid x^{k-1}) = W(i) \frac{P_i(x^{k-1})}{P(x^{k-1})}.
\end{align}
After the character $x_k$ is known, we can compare $\beta_i(k)$ and $\beta_i(k-1)$ and observe, that
\begin{align}
	\beta_i(k) = \beta_i(k-1) \frac{P_i(x_k\mid x^{k-1})}{f_k(x_k, P_1, P_2, \dots, P_m)} \text{ and } \beta_i(0) = W(i). \label{eq:beta_weighting_update}
\end{align}


\subsection{Generic Linear Weighting}

With the method of Lagrangian multipliers (see Section \ref{sec:geometric_model_divergence_minimization}) we can show that (in step $k$)
\begin{equation}
	P := \arg\min_{Q\in\P} \sum_{i=1}^m w_i D(P_i\parallel Q), \text{ where } w_i\geq0, 1\leq i\leq m, \text{ and } \sum_{i=1}^m w_i>0, \label{eq:linear_model_target1}
\end{equation}
yields the linear mixture
\begin{align}
	P(x\mid x^{k-1}) = f_k(x, P_1, P_2, \dots, P_m) := \sum_{i=1}^m w'_i P_i(x\mid x^{k-1}), \text{ where } w'_i := \frac{w_i}{\sum_{i=1}^m w_i}. \label{eq:generic_linear_mixture}
\end{align}
In the setting of the previous section the normalized weights $w'_i$ correspond to the switching probabilities $W(i\mid x^{k-1})$. Thus, the cost function in \eqref{eq:linear_model_target1} would be proportional to the expected redundancy of a switching source in step $k$.

It is important to understand the difference between \eqref{eq:geometric_model_target3} and \eqref{eq:linear_model_target1}. In \eqref{eq:geometric_model_target3} $P_i$ plays the role of a model distribution and we seek an approximate source distribution, which we can use as a model distribution. On the other hand, in \eqref{eq:linear_model_target1} $P_i$ plays the role of a source distribution and we seek a model distribution, which matches our \emph{assumptions} on the specific source structure (namely, a switching source). We belief that the assumptions of \eqref{eq:geometric_model_target3} are inferior to those of \eqref{eq:linear_model_target1}, hence the geometric mixture is more general.

In analogy to Section \ref{sec:geometric_model_weight_estimation} we look for a weight vector $\vect w^*$, which minimizes the code length of the sequence $x^n$ we want to compress, i.e.,
\begin{align}
	\vect w^* := \arg \min_{\vect w} \sum_{k=1}^n -\log \frac{\sum_{i=1}^m w_i P_i(x_k\mid x^{k-1})}{\sum_{i=1}^m w_i}. \label{eq:linear_model_generic_target1}	
\end{align}
First we analyse the convexity properties of \eqref{eq:linear_model_generic_target1}. W.l.o.g. we assume that $\vect w^T = (w_i)_{1\leq i\leq m}$ is an element of $\Omega_m$. The convexity properties of \eqref{eq:linear_model_generic_target1} follow from the analysis of a single term of the sum, which is proportional to
\begin{align}
	l(\vect w) := -\ln \frac{\vect w^T \vect P(x_k)}{\vect w^T \vect 1_m} \stackrel{\vect w\in\Omega_m}{=} \ln \frac{1}{\vect w^T \vect P(x_k)}, \text{ where } \vect P(x_k)^T := (P_i(x_k\mid x^{k-1}))_{1\leq i\leq m}.
\end{align}
The Hessian of $l(\vect w)$ is positive definite, since
\begin{align}
	\vect v^T \nabla^2 l(\vect w) \vect v = \vect v^T \frac{\vect P(x_k) \vect P(x_k)^T}{(\vect w^T \vect P(x_k))^2} \vect v = \left(\frac{\vect v^T \vect P(x_k)}{\vect w^T \vect P(x_k)}\right)^2 > 0
\end{align}
holds for $\vect v\neq \vect 0, ~\vect v\in\RR^m$. We conclude that the problem \eqref{eq:linear_model_generic_target1} is strictly convex. Thus, there exists a single global minimizer $\vect w^*\in\Omega_m$. As in Section \ref{sec:geometric_model_weight_estimation} we can obtain a weight update rule via iterative gradient descent
\begin{align}
	\vect w(k) := \max \left\{\varepsilon\vect 1_m, ~\vect w(k-1) + \alpha_k \frac{\vect P(x_k) - f_k(x_k, P_1, P_2, \dots, P_m) \cdot \vect 1_m}{f_k(x_k, P_1, P_2, \dots, P_m)\cdot \vect w^T \vect 1_m} \right\}, \label{eq:linear_model_generic_weight_update}
\end{align}
where $\vect w(0)^T := 1/m \cdot \vect 1_m$ and $\varepsilon$ is a small positive constant. It is interesting to note, that when we replace $\alpha_k$ with the matrix $\diag(\vect w(k-1))$ and omit the $\max$-operation, \eqref{eq:linear_model_generic_weight_update} turns into $\beta$-weighting (cf. \eqref{eq:beta_weighting_update}) and $\vect w(k)\in\Omega_m, k \geq 0$.


\section{Experiments} \label{sec:experimental_evaluation}
In this section we compare the performance of a geometric mixture ({\tt GEO}), a generic linear mixture ({\tt LIN}) and $\beta$-weighting ({\tt BETA}) on the files of the well-known Calgary Corpus. We have implemented the weighting techniques for a binary alphabet. To process non-binary symbols (here, bytes) we employ an alphabet decomposition. Every symbol $x_k\in\X$ is processed in $N = \lceil \log |\X|\rceil$ intermediate steps, for details see, e.g., \cite{cm_ccp2011}. To ensure a fair comparison, the set of models is the same for any mixture method: There are seven finite-order context models (the probability estimations are conditioned on order-$0$ to order-$6$ contexts). The eighth model is a \emph{match model}. In step $k$ it searches the longest matching substring $x^{k-1}_{k-L}$ of length $L\geq 7$ in $x^{k-2}$. In the case of a match it predicts the symbol (here, each bit in the $N$ intermediate steps), which succeeds the matching substring with probability $1 - 1/L$, otherwise each symbol receives the probability $1/|\X|$. 

For each mixture technique we select a weight vector $\vect w$ based on an order-$1$ context and on the match length $L$ (determined by the match model in every step $k$). Initially any weight vector is initialized to $1/m \cdot \vect 1_m$. After a weight update we ensure that $\vect w \geq \varepsilon \cdot \vect 1_m$ (we set $\varepsilon=2^{-30}$) and $\vect w^T \vect 1_m=1$. For $\beta$-weighting we can confirm the observation made in \cite{cm_cmidc}:
The weights must be bounded considerably away from zero, i.e., $\beta_i\geq \varepsilon$ (we set $\varepsilon=2^{-8}$).
A weight update based on iterative gradient descent requires a step size $\alpha_k$. We set $\alpha_k=1/16$ ({\tt GEO}) or $\alpha_k=1/32$ ({\tt LIN}), respectively. The step size (for {\tt GEO} and {\tt LIN}) and $\varepsilon$ (for {\tt BETA}) were  determined experimentally for maximum compression. We did not notice significant changes in compression, when the step size was sufficiently small (in the scale of $10^{-2}$).

Table \ref{tab:results} summarizes our experimental results. {\tt GEO} outperforms {\tt LIN} and {\tt BETA} in almost every case, expect for the file \textit{obj1}, where the compression is roughly $2\%$ worse than {\tt LIN} and {\tt BETA}. On average {\tt LIN} compresses about $2\%$ and {\tt BETA} compresses about $3.6\%$ worse than {\tt GEO}, respectively. When we compare {\tt LIN} and {\tt BETA} we see that {\tt BETA} produces worse compression in every case, $1.5\%$ on average. Summarizing we may say that {\tt GEO} works better than {\tt LIN}. In our experiments {\tt BETA} is inferior to the other weighting techniques.


\section{Conclusion}
In this paper we introduced geometric weighting as a new technique for computing mixtures in statistical data compression. In addition we introduced a new generic linear weighting strategy. We explain which assumptions the weighting techniques are based on. Furthermore, our results reveal that \ac{PAQ} is an instance of geometric weighting for a binary alphabet. All of the presented mixture techniques rely on weight vectors. It turns out that in any of the two cases the weight estimation is a good-natured problem since it is strictly convex. An experimental study indicates that geometric weighting is superior to linear weighting (for a binary alphabet).

For future research it would be interesting to obtain statements about the situations where geometric weighting outperforms linear weighting (and vice-versa). Another topic is how to select a fixed number of submodels for maximum compression. This leads to the optimization of model and mixture parameters (and to the question, whether or not, the optimization problem remains convex). Such a question is very natural, since we wish to maximize the compression with limited resources (CPU and RAM). Combining multiple models in data compression is highly successful in practice, but more research in this area is needed.

\textbf{Acknowledgment.} The author would like to thank Martin Dietzfelbinger, Michael Rink, Martin Aumueller and the anonymous reviewers for helpful comments and corrections.

\begin{table}[htbp]
	\caption{Compression rates in \ac{bpc} on the Calgary Corpus for geometric- ({\tt GEO}), generic linear- ({\tt LIN}) and $\beta$-weighting ({\tt BETA}), best results are typeset boldface.}
	\footnotesize{
	\begin{center}	
	\begin{tabular}{|c|c|c|c|}
	\hline
	\textbf{File}   & {\tt GEO} & {\tt LIN} & {\tt BETA} \\\hline 
	\textit{bib}    & \textbf{1.816} & 1.890 & 1.907 \\
	\textit{book1}  & \textbf{2.212} & 2.304 & 2.313 \\
	\textit{book2}  & \textbf{1.864} & 1.943 & 1.965 \\
	\textit{geo}    & \textbf{4.407} & 4.423 & 4.501 \\
	\textit{news}   & \textbf{2.286} & 2.347 & 2.412 \\
	\textit{obj1}   & 3.672 & \textbf{3.603} & 3.610 \\
	\textit{obj2}   & \textbf{2.224} & 2.240 & 2.298 \\
	\textit{paper1} & \textbf{2.274} & 2.327 & 2.343 \\
	\textit{paper2} & \textbf{2.220} & 2.288 & 2.310 \\
	\textit{pic}    & \textbf{0.813} & 0.871 & 0.922 \\
	\textit{progc}  & \textbf{2.276} & 2.327 & 2.361 \\
	\textit{progl}  & \textbf{1.558} & 1.607 & 1.651 \\
	\textit{progp}  & \textbf{1.610} & 1.638 & 1.669 \\
	\textit{trans}  & \textbf{1.384} & 1.430 & 1.453 \\\hline
	\textit{Average}& \textbf{2.187} & 2.231 & 2.265 \\\hline
	\end{tabular}
	\end{center}}	
	\label{tab:results}	
\end{table}

\vspace{-2mm}	
\bibliographystyle{plain}

\end{document}